\documentclass[12pt]{article}
\usepackage{epsfig}

\newcommand{\intl}{\int_0^\infty}
\newcommand{\eps}{\varepsilon}
\newcommand{\Li}{\mbox{\rm Li}}
\newcommand{\dalembertian}{\hbox{$\,\vbox{\hrule\hbox{\vrule%
  \vbox{\kern7pt}\kern7pt\vrule}\hrule}\,$}}

\begin{document} 

\begin{flushright}
MZ-TH/99-07\\
CLNS/99-1606\\
hep-ph/9903412\\
March 1999 \\
\end{flushright}
\vspace{0.5cm}
\begin{center}
{\Large\bf Configuration Space Based Recurrence Relations}\\[.5truecm]
{\Large\bf for Sunset-Type Diagrams}\\[1.3cm]
{\large S.~Groote$^{1,2}$, J.G.~K\"orner$^1$ and
  A.A.~Pivovarov$^{1,3}$}\\[1truecm]
$^1$ Institut f\"ur Physik, Johannes-Gutenberg-Universit\"at,\\[.2truecm]
Staudinger Weg 7, D-55099 Mainz, Germany\\[.5truecm]
$^2$ Floyd R.~Newman Laboratory of Nuclear Studies,\\[.2truecm]
Cornell University, Ithaca, NY 14853, USA\\[.5truecm]
$^3$ Institute for Nuclear Research of the\\[.2truecm]
  Russian Academy of Sciences, Moscow 117312
\vspace{1truecm}
\end{center}

\begin{abstract}
\noindent We derive recurrence relations for the calculation of multiloop
sunset-type diagrams with large powers of massive propagators. The technique
is formulated in configuration space and exploits the explicit form of the
massive propagator raised to a given power. We write down and evaluate a
convenient set of basis integrals. The method is well suited for a numerical
evaluation of this class of diagrams. We give explicit analytical formulae
for the basis integrals in the asymptotic regime. 
\end{abstract}

\newpage

\section{Introduction}
With the present accuracy of phenomenological applications, high precision
tests of the Standard Model and search for new physics invariably require
accounting for higher order contributions in coupling constants within
perturbation theory. This necessitates the computation of multiloop
diagrams~\cite{SMrev}. Numerous high precision tests of the Standard Model
require the evaluation of diagrams in the approximation when external
momenta are small in comparison with the masses of particles corresponding
to internal lines of the diagrams (see e.g.~\cite{rho,rhop,ChPhysRep}). 
In this way massive diagrams without external momenta -- vacuum bubbles
-- appear.

The dominant technique for high-order perturbation theory calculations of
multiloop diagrams nowadays consists in using recurrence relations obtained
from the method of integration by part within dimensional
regularization~\cite{ibyparts}. The final answer of a calculation is
expressed in terms of a few master integrals that serve as initial values
for the recurrence equations. These master integrals are explicitly
evaluated (analytically or numerically) while the reduction of a given
diagram to the master integrals involves only algebraic manipulations (as
an example see ref.~\cite{b3b4} which presents the idea in its full bright
glory).

Presently the evaluation of massive vacuum bubbles diagrams at the
three-loop level is basically completed. The general strategy of reducing 
all massive three-loop bubbles to a basic set of master integrals through
recurrence relations was described in ref.~\cite{avdeev}. Recently the
remaining unknown master integrals have been analytically identified by
using high precision numerical computation assuming a given basis of
transcendental numbers for master diagrams~\cite{broad}. The existing
techniques for the three-loop evaluation of massive vacuum bubbles make use
of the reduction of six-dimensional objects (in the parameter space of
powers of the propagators) through recurrence relations obtained within the
integration by part technique. In direct applications of these techniques
some topology classes of diagrams (sunset-type topology with large powers
of propagators) appear to be one of the main sources of heavy computer time
consumption~\cite{chetpriv}. Nevertheless, in principle, the problem of
analytical calculation of three-loop bubbles is solved.

However, in practice the solution to the recurrence relations is not a
routine procedure and requires the manipulation of a huge number of terms.
Even using symbolic manipulation programs this task is in some cases
beyond present computer capabilities. For instance, the computation of
higher moments of the $b$-quark spectral function~\cite{ChKSt1,ChKSt2},  
which is important for a precision determination of the $b$-quark
mass~\cite{KPP}, is limited by these obstacles and only moments up to
$n=8$ are presently available within perturbation theory~\cite{ChKStmom8}.
In view of the importance of this problem new ideas and techniques to
improve on present results are called for.

Some attention has recently been drawn to the problem of optimizing the
recurrence procedure for three-loop bubbles to find shorter routes to the
final solution (see e.g.~\cite{baikov} and refs.\ therein). It has been
proven that the final result for three-loop bubbles contains only several
transcendental numbers (which are known) with rational coefficients for
arbitrary mass configuration~\cite{broad}. This structure of the final answer
prompts one to search for more direct ways of obtaining the physical results.
Indeed, for phenomenological applications one only needs numerical values 
of the coefficients of perturbation theory and their analytical expressions
are not really important~\cite{rho,rhop}. The advantage of analytical
computation is, of course, the full control over precision. If numerical
methods are used, one has to take extra care about possible error
accumulation. This is not always straightforward in this type of calculation 
because huge numerical cancellations may occur among the large number
of contributing terms.

We discuss a technique that allows one to reduce these diagrams to a
one-parameter set of basis integrals of a rather simple structure. The
method is formulated in configuration space and exploits the explicit form
of massive propagators with large powers of denominators. A convenient
set of basis integrals is written down. The generalization to any number
of loops is straightforward which is a part of the motivation for the
present investigation. The method is well suited for a numerical evaluation
of this class of diagrams.

In the course of comparing our analytical results with those given in
ref.~\cite{b3b4} we have derived a number of interesting new analytical
results on definite integrals involving products of McDonald functions with
powers and logarithms. 

The paper is organized as follows. To begin with, in Sec.~2 we remind the
reader of some features of the configuration space approach and introduce
our notation which closely follows the notation in ref.~\cite{wm}. In
Sec.~3 we present several explicit examples taken from the subclass of
sunset-type diagrams, the water melon diagrams, in order to compare the
results obtained within the configuration space approach with results given
in the literature~\cite{b3b4}. In Sec.~4 we explain our general recursion
concept and present a convenient set of basis integrals. In Sec.~5 we
present different methods to calculate these basis integrals. Our
conclusions are given in Sec.~6.

\section{Basic relations}
We consider three-loop vacuum bubbles with only one mass $m$ which serves
as a dimensional parameter. The classification of the topology prototypes 
for three-loop vacuum bubbles was presented in ref.~\cite{avdeev}. The
analytical computation of some missing master integrals has recently been
completed~\cite{broad}. However, the solution of the recurrence relations 
leading to the master integrals is complicated and time consuming,
especially for large powers of propagators. We suggest new recurrence
relations for a particular topology of vacuum bubbles which allows for an
explicit solution. The simplicity of our technique is manifest in the 
configuration space representation for Feynman diagrams. First we remind
the reader of some features of the configuration space approach and
introduce our notation which closely follows the notation in ref.~\cite{wm}.

\begin{figure}\begin{center}
\epsfig{figure=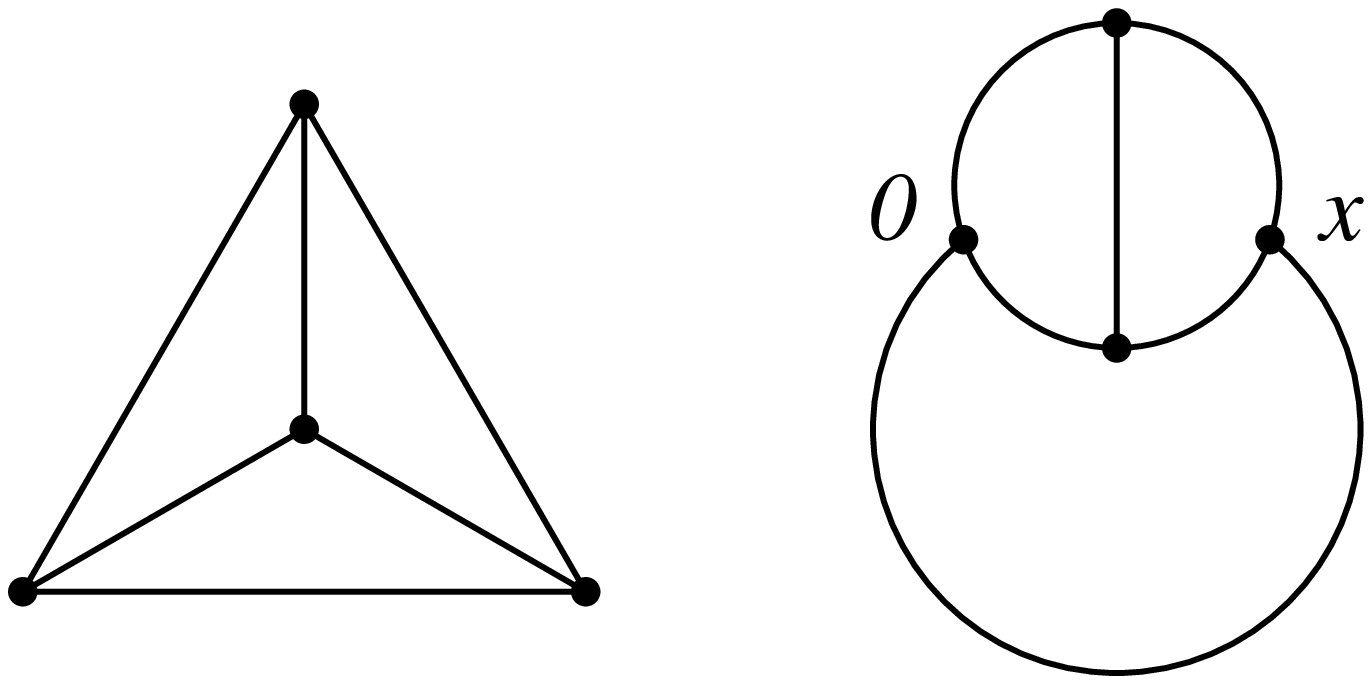, height=5truecm, width=10truecm}\vspace{12pt}
\hbox{\bf\kern3truecm(a)\kern6truecm(b)\kern3truecm}
\caption{\label{fig1}Three-loop vacuum bubble diagram in two different
representations: (a) the tetrahedron representation, (b) the
``fish+propagator'' representation where the configuration space points $0$
and $x$ are indicated.}
\end{center}\end{figure}

A general three-loop vacuum diagram has the topology of a tetrahedron. It
also can be thought of as ``fish+propagator'' topology (see Fig.~1) where
the ``fish'' part is a master two-loop diagram. This topology
suggests the use of the following representation in configuration space
(with an obvious choice of space-time points)
\begin{equation}
\Pi(x)={\rm fish\/}(x)D(x,m)
\label{represent}
\end{equation}
where $D(x,m)$ is a propagator of a massive particle with mass $m$ in
$D$-dimensional (Euclidean) space-time,
\begin{equation}\label{eqn02}
  D(x,m)=\frac1{(2\pi)^D}\int\frac{e^{-ip_\mu x^\mu}d^Dp}{p^2+m^2}
  =\frac{(mx)^\lambda K_\lambda(mx)}{(2\pi)^{\lambda+1}x^{2\lambda}},\qquad
  D(x,0)=\frac{\Gamma(\lambda)}{4\pi^{\lambda+1}x^{2\lambda}}.
\end{equation}
We write $D=2\lambda+2$, $\lambda=1-\eps$; $K_\lambda(z)$ is the McDonald
function (a modified Bessel function of the third kind, see
e.g.~\cite{Watson,GrRy}). The propagator $D(x,m)$ depends only on the
length of the space-time vector $|x|=\sqrt{x_\mu x^\mu}$ for which we
simply write $x$. The explicit representation for the modified internal
line with mass $m$ (power of the massive propagator) is given by
\begin{equation}
D^{(\mu)}(x,m)=\frac1{(2\pi)^D}\int
  \frac{e^{-ip_\nu x^\nu}d^Dp}{(p^2+m^2)^{\mu+1}}
  =\frac1{(2\pi)^{\lambda+1}2^\mu\Gamma(\mu+1)}
  \left(\frac mx\right)^{\lambda-\mu}K_{\lambda-\mu}(mx).
\label{modifprop}
\end{equation}
It contains the same functions (up to the difference in indices) as
Eq.~(\ref{eqn02}) and thus does not change the general functional structure
of the representation constructed below. This is the reason why our method
is well suited for dealing with large powers of propagators of massive
particles.

The quantity of interest -- a vacuum bubble -- is represented by the
integral
\begin{equation}
\tilde\Pi(0)=\int\Pi(x)d^Dx
\label{Fourier}
\end{equation}
which is nothing but the Fourier transform of $\Pi(x)$ at zero momentum. It
depends on a single dimensional parameter $m$, the mass of the particles in
the massive lines. 

The ``fish'' part of the diagram is written in terms of a dispersion
relation in configuration space, 
\begin{equation}
{\rm fish\/}(x)=\int \rho_f(s)D(x,\sqrt{s})ds
\label{disp}
\end{equation}
which leads to a representation of the form
\begin{equation}
\tilde\Pi(0)=\int\Pi(x)d^Dx=\int ds\,\rho_f(s)\int D(x,\sqrt s)D(x,m)d^Dx .
\label{final}
\end{equation}
This form can be further simplified by performing the integration in $x$
explicitly. The result reads
\begin{eqnarray}
\lefteqn{\int D(x,\sqrt s)D(x,m)d^Dx
  \ =\ \left(\frac{m\sqrt s}2\right)^\lambda
  \frac1{(2\pi)^{\lambda+1}\Gamma(\lambda+1)}
  \intl x K_\lambda(x\sqrt{s})K_\lambda(mx)dx}\nonumber\\
  &=&\left(\frac{m\sqrt s}2\right)^\lambda
  \frac\pi{(2\pi)^{\lambda+1}\Gamma(\lambda+1)}
  \Gamma\left(\frac12+\lambda\right)\Gamma\left(\frac12-\lambda\right)
  {}_2 F_1\left(\frac12+\lambda,\frac12;1;1-\frac{m^2}{s}\right).\nonumber
\end{eqnarray}
Here ${}_2 F_1(a,b;c;z)$ is a hypergeometric function. Therefore the
problem of evaluating the diagram is reduced to a one-dimensional integral
if the spectral density for the ``fish'' part $\rho_f(s)$ is known. The
spectral density $\rho_f(s)$ was computed for the principal massive
configurations in four-dimensional space-time~\cite{spectrbroad}. If the
propagator that multiplies the ``fish'' part of the diagram is massless,
this formula can be further simplified and is given by 
\[
\int D(x,\sqrt{s})D(x,0)d^Dx
  =\frac{s^{\lambda/2}}{2\lambda(2\pi)^{\lambda+1}}
  \intl x^{1-\lambda} K_\lambda(x\sqrt{s})dx
  =\left({s\over 2}\right)^{\lambda/2}
\frac{\Gamma(1-\lambda)}{4\lambda(2\pi)^{\lambda+1}}.
\]
The final integration over $s$ in Eq.~(\ref{final}) now includes only some
power of the energy square $s$ (instead of the hypergeometric function
as in the massive case) and is rather straightforward. The equivalent set of
formulae can also be obtained in the momentum space
representation~\cite{broad}.

\begin{figure}\begin{center}
\epsfig{figure=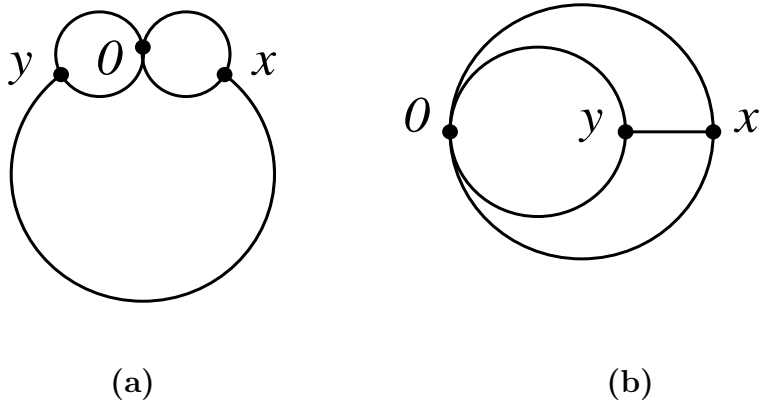, height=4truecm, width=10truecm}\vspace{12pt}
\hbox{\bf\kern3truecm(a)\kern6truecm(b)\kern3truecm}
\caption{\label{fig2}The  ``spectacle+propagator'' representation, also called 
  ``spectacle'' topology diagram in two different forms, namely
  (a) the form used in Eq.~(\ref{spectxexp}) and
  (b) the form used in Eq.~(\ref{midline}).}
\end{center}\end{figure}

In some cases one propagator can be removed from a diagram using the
recurrence relations for bubbles obtained within the integration
by part technique~\cite{avdeev}. Then the diagram becomes simpler. The initial
``fish+propagator'' topology converts to a ``spectacles+propagator''
topology. The spectral density for the spectacle part can be computed in a
rather simple way. It is given by a product of two one-loop integrals in
the momentum space representation (see Fig.~2). In the original
classification of~\cite{avdeev} these are class $E$ diagrams. We will
henceforth adopt the classification of~\cite{avdeev} to denote the different
topology classes of diagrams.

\begin{figure}\begin{center}
\epsfig{figure=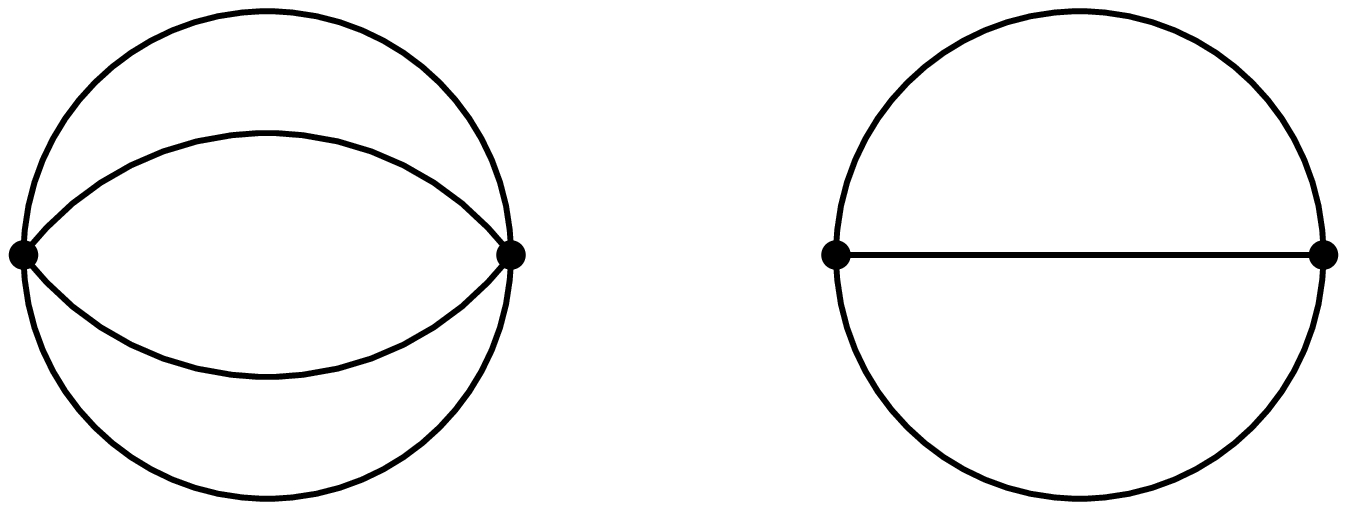, height=4truecm, width=10truecm}\vspace{12pt}
\hbox{\bf\kern3truecm(a)\kern6truecm(b)\kern3truecm}
\caption{\label{fig3}(a) Three-loop (i.e.\ four-line) water melon and
(b) two-loop water melon (three-line water melon, being the ordinary sunset
diagram).}
\end{center}\end{figure}

After a deliberate use of the recurrence relations for bubbles, in some
cases two propagators can be removed. A typical situation of such a
kind was analyzed in ref.~\cite{b3b4}. In such a case the diagrams
become simple indeed. Even the most complicated ones belong to the subclass
of water melon topologies which can be computed immediately (see Fig.~3).
Their properties (with generalization to any number of loops) have been
described in detail in ref.~\cite{wm}. The most attractive feature of such
a strategy is that for high derivatives of propagators (large powers of
denominators) the corresponding recurrence relations for this particular
topology can be solved very efficiently.

\section{Water melon topology: explicit examples}
While the water melon class of diagrams can appear as part of the remnants
of the general recursive procedure, there are some cases when they are just
the final aim of the recursion procedure. This is the case for the $B_N$
subclass of diagrams~\cite{b3b4}. Some of the master integrals (for
instance, $D_3(0,1,0,1,1,1)$ in ref.~\cite{avdeev}) are exactly water melons.
A further simplification of water melon diagrams can be achieved with the
use of their particular properties. In the configuration space
representation the water melon diagrams can be reduced to a specific basis
set of simple integrals quite efficiently. 

The configuration space technique for the water melon topology is
particularly convenient for a numerical evaluation since it is simple and
efficient. We have reproduced numerically some results of
refs.~\cite{b3b4,broad} with our technique.

In the particular case of the subclass $B_N$ of the bubble diagrams, the
water melon topology diagrams emerge naturally and can be chosen as master
configurations. The reduction of a general diagram of this subclass to the
water melon topology is explicitly constructed in ref.~\cite{b3b4}. Here we
discuss a representation which can be obtained within the configuration space
technique. The representation is simple and is given by a one-dimensional 
integral. The expansion in $\eps$ within dimensional regularization is
straightforward and is explicitly given for the evaluation of the numerical
value of the integral $B_4$~\cite{b3b4} for which we give a new
representation.

The starting point of our calculation is the definition of the $B_N$ class
of diagrams~\cite{b3b4},
\begin{eqnarray}
\lefteqn{B_N(0,0,n_3,n_4,n_5,n_6)\ =\ \int\frac{d^Dk\,d^Dl\,d^Dp}
  {m^{3D}(\pi^{D/2}\Gamma(1+\eps))^3}\ \times}\nonumber\\&&
  \frac{m^{2n_3}}{((p+k)^2+m^2)^{n_3}}\frac{m^{2n_4}}{((p+l)^2+m^2)^{n_4}}
  \frac{m^{2n_5}}{((p+k+l)^2+m^2)^{n_5}}\frac{m^{2n_6}}{(p^2+m^2)^{n_6}}
\end{eqnarray}
with two propagators absent ($n_1=n_2=0$) to obtain a water melon topology
for the three-loop case. In the following we will suppress these first two
indices in the notation for the $B_N$ class diagrams. The configuration
space expression for the generalized propagator (with crosses or having
differentiated in its mass or momentum) is given by Eq.~(\ref{modifprop}).
This can be inserted into the above expression for $B_N$ and after
rearrangement of integrations leads to
\begin{eqnarray}
\lefteqn{B_N(n_3,n_4,n_5,n_6)\ =\
  \frac{m^{2(n_3+n_4+n_5+n_6)-3D}}{(\pi^{D/2}\Gamma(1+\eps))^3}\ \times}\\&&
  (2\pi)^{3D}\int D^{(n_3-1)}(x,m)D^{(n_4-1)}(x,m)
  D^{(n_5-1)}(x,m)D^{(n_6-1)}(x,m)d^Dx\nonumber
\end{eqnarray}
which can be reduced to a one-dimensional integral using the rotational
invariance of the integration measure in Euclidean space-time,
\begin{equation}
d^Dx=\frac{2\pi^{\lambda+1}}{\Gamma(\lambda+1)} x^{2\lambda+1}dx.
\end{equation}
Note that various techniques of eliminating tensorial structures for vacuum
diagrams were discussed in refs.~\cite{wm,ChetTens}. On the other hand
we have~\cite{wm}
\begin{eqnarray}
\tilde \Pi(0)&=&\frac{2\pi^{\lambda+1}}{\Gamma(\lambda+1)}\intl
  D^{(n_3-1)}(x,m)D^{(n_4-1)}(x,m)\ \times\nonumber\\&&\qquad
  D^{(n_5-1)}(x,m)D^{(n_6-1)}(x,m)x^{2\lambda+1}dx.
\label{main0}
\end{eqnarray}
The comparison of these two formulas results in
\begin{equation}
B_N(n_3,n_4,n_5,n_6)=\frac{(2\pi)^{3D}m^{2(n_3+n_4+n_5+n_6)-3D}}
  {(\pi^{D/2}\Gamma(1+\eps))^3}\tilde\Pi(0)
\label{main}
\end{equation}
where the powers of the propagators in $\tilde\Pi(0)$ have been
appropriately adjusted. In the following calculations we set $m=1$ for the
mass (regardless of the units, of course). We will discuss some explicit
applications of Eqs.~(\ref{main0}) and~(\ref{main}) in the following to
indicate their properties.

As a first example we consider the integral $B_N(2,2,2,2)$ in the case
$\lambda=1$ (four-dimensional space-time). Generally we have the explicit
expressions 
\begin{equation}
D^{(1)}(x,1)=\frac{x^{1-\lambda}}{(2\pi)^{\lambda+1}2^1\Gamma(2)}
  K_{\lambda-1}(x)=\frac{x^\eps}{2(2\pi)^{2-\eps}}K_{-\eps}(x)
\end{equation}
for the propagators which results in
\begin{equation}
D^{(1)}(x,1)=\frac1{2(2\pi)^2}K_0(x)
\end{equation}
for $\eps=0$. We obtain
\begin{eqnarray}
B_N(2,2,2,2)=\frac{(2\pi)^{12}}{\pi^6}
  \frac{2\pi^2}{16(2\pi)^8}\intl K_0^4(x)x^3dx
  =2\intl K_0^4(x)x^3dx
\end{eqnarray}
which is the expression obtained earlier in~\cite{wm}. Note that the
function $K_0(x)$ is a propagator of a massive particle in two-dimensional
space-time. Therefore many results can be obtained by using two-dimensional
field theory in Euclidean space-time (see e.g.\ ref.~\cite{schPL}).

For the more general case of $D$-dimensional space-time we obtain
\[
\tilde\Pi(0)=\frac{2\pi^{2-\eps}}{16(2\pi)^{8-4\eps}\Gamma(2-\eps)}
  \intl x^{4\eps}K_{-\eps}^4(x)x^{3-2\eps}dx\nonumber 
\]
and 
\begin{equation}
B_N(2,2,2,2)=\frac{2^{1-2\eps}}{(1-\eps)\Gamma(1+\eps)^3\Gamma(1-\eps)}
  \intl K_{-\eps}^4(x)x^{3+2\eps}dx.
\label{epex}
\end{equation}
To find higher orders in the $\eps$-expansion necessary for computations
within dimensional regularization we use series expansions in $\eps$ of all
quantities entering Eq.~(\ref{epex}). First we have 
the rather obvious results
\begin{eqnarray}
\frac{2^{1-2\eps}}{(1-\eps)\Gamma(1+\eps)^3\Gamma(1-\eps)}
  &=&2(1+\eps-2\eps\ln 2+2\eps\gamma_E)+O(\eps^2),\nonumber\\[2pt]
x^{3+2\eps}&=&x^3(1+2\eps\ln x)+O(\eps^2),
\end{eqnarray}
where $\gamma_E$ is Euler's constant. Within the dimensional regularization
scheme the propagator in the configuration space contains the McDonald
function with a non-integer index depending on the regularization parameter
$\eps$. To expand the McDonald function in the parameter $\eps$ entering
its index, we use the general formula~\cite{GrRy}
\begin{equation}
\left[\frac{\partial K_\nu(z)}{\partial\nu}\right]_{\nu=\pm n}
  =\pm\frac12n!\sum_{k=0}^{n-1}\left(\frac z2\right)^{k-n}
  \frac{K_k(z)}{k!(n-k)},\qquad n\in\{0,1,\ldots\,\}
\label{eexpex}
\end{equation}
for the derivative of the McDonald function with respect to its index near 
integer values of this index. In this case one obtains
\begin{equation}
K_{-\eps}(x)=K_0(x)+O(\eps^2).
\label{Kexp}
\end{equation}
We end up with
\begin{eqnarray}
B_N(2,2,2,2)&=&2\intl K_0^4(x)x^3dx+2\eps(1+2\ln 2+2\gamma_E)
  \intl K_0^4(x)x^3dx\nonumber\\&&\qquad
  +4\eps\intl K_0^4(x)x^3\ln x\,dx+O(\eps^2)\nonumber\\
  &=&2I_0(3)+2\eps(1-2\ln 2+2\gamma_E)I_0(3)+4\eps I_0^l(3)+O(\eps^2)
\label{e2222}
\end{eqnarray}
where a general notation for the configuration space integrals
\begin{eqnarray}
I_m(q)&=&\intl K_0^{4-m}(x)K_1^m(x)x^qdx,\nonumber\\
I_m^l(q)&=&\intl K_0^{4-m}(x)K_1^m(x)x^q\ln x\,dx
\label{defint}
\end{eqnarray}
is introduced. The only new contribution in the $\eps$-expansion till the
first order in Eq.~(\ref{e2222}) is connected with the logarithmic integral
$I_0^l(3) ($$I_m^l(q)$ from Eq.~(\ref{defint})). The part related to
$I_0(3)$ in this order is a trivial kinematic contribution. The term 
$2(\ln 2-\gamma_E)$ in Eq.~(\ref{e2222}) can be easily removed by
redefining the logarithmic integral using $\ln x\to\ln(xe^{\gamma_E}/2)$.

Identifying the parameters $B_3$ and $B_4$ from~\cite{b3b4} we find
\begin{eqnarray}
\label{ident}
B_N(2,2,2,2)&=&-\frac38+\frac7{16}B_3
  +\left(\frac{63}{32}B_3+\frac3{16}B_4\right)\eps+O(\eps^2)\\
  &=&-\frac38+\frac7{16}\zeta(3)+\left(\frac{63}{32}\zeta(3)
  -\frac{63}{32}\zeta(4)+\frac3{16}B_4\right)\eps+O(\eps^2)\nonumber
\end{eqnarray}
where $B_3=\zeta(3)-\frac92\eps\zeta(4)+O(\eps^2)$. The comparison of the
zeroth order term of Eq.~(\ref{ident}) with Eq.~(\ref{e2222}) results in 
the relation 
\begin{equation}
I_0(3)=-\frac3{16}+\frac7{32}\zeta(3)
\end{equation}
which assigns a value to one of the initial terms of the recurrence relations
that will be presented later in Sec.~4. We checked this relation
numerically. In the first order part of the $\eps$-expansion we solve for
$B_4$ obtaining the representation
\begin{eqnarray}
\label{firstb4}
B_4&=&\frac{16}3\Big(2(1-2\ln 2+2\gamma_E)I_0(3)+4I_0^l(3)+\frac{63}{32}
  (\zeta(4)-\zeta(3))\Big)\nonumber\\
  &=&\frac{32}3\Big((1-2\ln 2+2\gamma_E)I_0(3)+2I_0^l(3)\Big)
  +\frac{21}2\Big(\zeta(4)-\zeta(3)\Big)
\end{eqnarray}
which after substituting for $I_0^l(3)$ from Eq.~(\ref{defint}) gives
numerically $B_4=-1.7628\ldots$ This numerical value expressed in terms of
configuration space integrals within our technique coincides with the
result given in~\cite{b3b4}. Taking the analytical expression for $B_4$
from~\cite{b3b4}
\begin{equation}
B_4=16{\rm Li}_4\left(\frac12\right)+\frac23\ln^4(2) 
  -\frac23\pi^2\ln^2(2)-\frac{13}{180}\pi^4 
\label{b4an}
\end{equation}
with ${\rm Li}_4(z)$ being a fourth order polylogarithm, 
\[
{\rm Li}_4(z)=\sum_{k=1}^\infty\frac{z^k}{k^4},\qquad|z|<1,
\]
we obtain the result for the logarithmic integral $\tilde I_0^l(3)$,
\begin{eqnarray}
\tilde I_0^l(3)&=&\intl K_0^4(x)\ln(xe^{\gamma_E}/2)x^3\,dx\\
&=&\frac3{32}+\frac34{\rm Li}_4\left(\frac12\right)-\frac{17}{1920}\pi^4 
  -\frac1{32}\pi^2\ln^2(2)+\frac1{32}\ln^4(2)+\frac{49}{128}\zeta(3)\nonumber 
\label{b4num}
\end{eqnarray}
which serves as the initial value for the recurrence relations for the set
of logarithmic integrals (Sec.~4). We have checked this value numerically.
For a further demonstration of the efficiency of the configuration space
technique for fixed powers of the propagators we calculate the two
integrals $B_N(2,2,2,1)$ and $B_N(2,3,3,4)$ (the latter does not contain
$B_4$ which is the reason for having selected this example).

For the integral $B_N(2,2,2,1)$ in four-dimensional space-time ($\lambda=1$)
we have to include the propagator
\begin{equation}
D^{(0)}(x,1)=\frac{x^{-\lambda}}{(2\pi)^{\lambda+1}2^0\Gamma(1)}
  K_\lambda(x)=\frac{x^{\eps-1}}{(2\pi)^{2-\eps}}K_{1-\eps}(x)
\end{equation}
equal to $K_1(x)/4\pi^2x$ for $\lambda=1$ which is a standard propagator of
a massive particle for $D=4$. We obtain a representation of the form 
\begin{equation}
B_N(2,2,2,1)=4\int K_0^3(x)K_1(x)x^2dx=4I_1(2).
\label{e2221}
\end{equation}
For the integral $B_N(2,2,2,1)$ in the case $\lambda=1-\eps$ we obtain the
generalization of Eq.~(\ref{e2221}) of the form
\begin{eqnarray}
B_N(2,2,2,1)=\frac{2^{2-2\eps}}{(1-\eps)\Gamma(1+\eps)^3\Gamma(1-\eps)}
  \intl K_{-\eps}^3(x)K_{1-\eps}(x)x^{2+2\eps}dx.
\end{eqnarray}
The $\eps$-expansion of the factor multiplying the integral is the same as
in the former case except for an overall factor of $2$. A similar statement
is valid for the expansion of the power of $x$. What remains to be done is
to expand the McDonald functions in the vicinity of integer values of their
indices. To obtain this expansion we use the relation
\begin{equation}
\frac{\partial K_\nu(x)}{\partial\nu}\Big|_{\nu=1}
  =\frac12\left(\frac x2\right)^{-1}K_0(x)=\frac1xK_0(x)
\end{equation}
which contributes to the power expansion as
\begin{equation}
K_{1-\eps}(x)=K_1(x)-\frac\eps xK_0(x)+O(\eps^2).
\end{equation}
Using these expansions we obtain the representation in terms of our basic
integrals
\begin{eqnarray}
B_N(2,2,2,1)&=&4\intl x^2K_0^3(x)K_1(x)dx\nonumber\\&&
  +4\eps(1-2\ln 2+2\gamma_E)\intl x^2K_0^3(x)K_1(x)dx\nonumber\\&&
  -4\eps\intl xK_0^4(x)dx+8\eps\intl x^2\ln x\,K_0^3(x)K_1(x)dx
  +O(\eps^2)\nonumber\\
  &=&4I_1(2)+4\eps(1-2\ln 2+2\gamma_E)I_1(2)-4\eps I_0(1)+8\eps I_1^l(2)
  +O(\eps^2)\nonumber
\end{eqnarray}
that has to be compared with the output of the RECURSOR package~\cite{b3b4}
with the explicit master integrals $B_3$, $B_4$,
\begin{eqnarray}
B_N(2,2,2,1)&=&\frac74B_3+\frac34B_4\eps+O(\eps^2)\nonumber\\
&=&\frac74\zeta(3)+\left(\frac34B_4-\frac{63}8\zeta(4)\right)\eps+O(\eps^2).
\end{eqnarray}
The zeroth order comparison gives the result $I_1(2)=\frac7{16}\zeta(3)$
which has been verified numerically. The first order comparison results in
\begin{eqnarray}
B_4&=&\frac43\big(4(1-2\ln 2+2\gamma_E)I_1(2)-4I_0(1)+8I_1^l(2)
  +\frac{63}8\zeta(4)\big)\\
  &=&\frac{16}3\big((1-2\ln 2+2\gamma_E)I_1(2)-I_0(1)+2I_1^l(2)\big)
  +\frac{21}2\zeta(4)\ =\ -1.7628\nonumber
\end{eqnarray}
as before (see Eq.~(\ref{firstb4})). For the integral $B_N(2,3,3,4)$ in the
case $\lambda=1-\eps$ we need to include further propagators. They are
\begin{eqnarray*}
D^{(2)}(x,1)&=&\frac{x^{2-\lambda}}{(2\pi)^{\lambda+1}2^2\Gamma(3)}
  K_{\lambda-2}(x)\ =\ \frac{x^{1+\eps}}{8(2\pi)^{2-\eps}}K_{-1-\eps}(x),\\
D^{(3)}(x,1)&=&\frac{x^{3-\lambda}}{(2\pi)^{\lambda+1}2^3\Gamma(4)}
  K_{\lambda-3}(x)\ =\ \frac{x^{2+\eps}}{48(2\pi)^{2-\eps}}K_{-2-\eps}(x).
\end{eqnarray*}
Both McDonald functions have to be expanded in their index. We have 
$K_{-1}(x)=K_1(x)$ and
\[
\frac{\partial K_\nu(x)}{\partial\nu}\Big|_{\nu=-1}
  =-\frac12\left(\frac x2\right)^{-1}K_0(x)=-\frac1xK_0(x),
\]
thus
\[
K_{-1-\eps}(x)=K_1(x)+\frac\eps xK_0(x)+O(\eps^2),
\]
and
\begin{eqnarray*}
\frac{\partial K_\nu(x)}{\partial\nu}\Big|_{\nu=-2}
  &=&-\frac122!\left(\frac1{2!}\left(\frac x2\right)^{-2}K_0(x)
  +\frac1{1!1!}\left(\frac x2\right)^{-1}K_1(x)\right)\nonumber\\
  &=&-\frac2{x^2}K_0(x)-\frac2xK_1(x),
\end{eqnarray*}
so that
\begin{equation}
K_{-2-\eps}=K_2(x)+\frac{2\eps}xK_1(x)+\frac{2\eps}{x^2}K_0(x)+O(\eps^2).
\end{equation}
With these relations we obtain the following $\eps$-expansion for the
integral in question
\begin{eqnarray}
B_N(2,3,3,4)&=&\frac{2^{-2\eps}}{192(1-\eps)\Gamma(1+\eps)^3\Gamma(1-\eps)}
  \ \times\nonumber\\&&\qquad
  \intl K_{-\eps}(x)K_{-1-\eps}^2(x)K_{-2-\eps}(x) x^{7+2\eps}dx\nonumber\\
  &=&\frac1{192}I_{21}(7)+\frac\eps{192}(1-2\ln 2+2\gamma_E)I_{21}(7)
  +\frac\eps{96}I_{11}(6)\nonumber\\&&
  +\frac\eps{96}I_3(6)+\frac\eps{96}I_2(5)
  +\frac\eps{96}I_{21}^l(7)+O(\eps^2)
\end{eqnarray}
where we have introduced the generalized integral
\begin{equation}
I_{mn}(q)=\intl K_0^{4-m-n}(x)K_1^m(x)K_2^n(x)x^qdx.
\end{equation}
This expansion has to be compared with the representation through master
integrals found in momentum space,
\begin{equation}
B_N(2,3,3,4)=\frac1{576}+\left(\frac{385}{65536}B_3
  -\frac{809}{884736}\right)\eps+O(\eps^2)
\end{equation}
resulting in the identification
\begin{equation}
I_{21}(7)=\intl K_0(x)K_1^2(x)K_2(x)x^7dx=\frac13
\end{equation}
which is surprisingly simple and contains no transcendental numbers usually
present in such integrals. It is rather curious that a similar 
identification allows one to express $\zeta(3)$ in terms of our basis integrals
\begin{eqnarray}
\zeta(3)&=&\frac{1024}{1155}\Big((1-2\ln 2+2\gamma_E)I_{21}(7)+2I_{11}(6)
  \nonumber\\&&
  +2I_3(6)+2I_2(5)+2I_{21}^l(7)+\frac{809}{4608}\Big) .
\end{eqnarray}
We checked both equations numerically to make certain that they are valid.
These results serve as a hint that the standard basis may not be the
simplest and most relevant basis to be used in
computations of massive three-loop diagrams.

We have considered some explicit expressions following from the
configuration space representation of water melon diagrams. In all these
examples the integrals are ultraviolet finite. We emphasize that
ultraviolet divergences add nothing new to the analysis. For this
particular topology the structure of the ultraviolet divergences is
particularly simple: all divergences result from the region of
integration around small $x$ and there are no overlapping divergences.
Therefore the divergent parts of the diagrams can be obtained by expanding
the integrand at small $x$ and subtracting the corresponding singularities
which simplifies the integrands drastically (making them effectively
massless) and the integration of the divergent parts can be done
analytically~\cite{wm} within some (usually dimensional) regularization
scheme. More complicated is to find the finite parts of the massive
diagrams. Keeping this in mind we do not discuss the problem of ultraviolet
divergences in the rest of the paper and concentrate on the finite parts. 

\section{Reduction}
The preceding section has shown that three-loop water melon diagrams can be
expressed as configuration space integrals of a product of at most four
McDonald functions $K_\nu(x)$ where $\nu$ need not be an integer. In this
section we want to exhibit the three steps of how to reduce the set of
necessary integrals to a smaller set. First we can get rid of the non-integer
dimensionality of the functions by using Eq.~(\ref{eexpex}) to expand into
powers of $\eps$ resulting in integrals containing a product of four
McDonald functions with or without a factor $\ln(x)$. As a second step we
use the relation
\begin{equation}
K_n(x)=2\frac{n-1}xK_{n-1}(x)+K_{n-2}(x)
\label{mainrecur}
\end{equation}
for McDonald functions of different orders to further reduce the integrals
to integrals only containing $K_0^4$, $K_0^3K_1$, $K_0^2K_1^2$, 
$K_0K_1^3$ and $K_1^4$ together with some positive powers of $x$ and again
with or without a factor $\ln(x)$. The last step consists in using
\begin{equation}
\frac{d}{dx}K_\nu(x)=-\frac{1}{2}(K_{\nu-1}(x)+K_{\nu+1}(x))
\end{equation}
for integer $\nu$ (which is valid for any complex $\nu$ as well) and a
partial integration in order to reduce the necessary integrals to integrals
containing only the McDonald function $K_0(x)$. This reduction procedure  
will be considered for two different cases in the following two subsections.

\subsection{Reduction to integrals containing $K^4_0(x)$}
After using the recurrence relations for Bessel functions the general
water melon diagram reduces to a linear combination of basic integrals
given in Eq.~(\ref{defint}). The subsequent reduction to even simpler
integrals containing only $K^4_0(x)$ and powers of $x$ is done by partial
integration. For this purpose we make use of the relations
\begin{eqnarray}
\frac{d}{dx}K_0(x)&=&-K_1(x)\qquad\mbox{and}\nonumber\\
\frac{d}{dx}K_1(x)&=&-\frac12(K_0(x)+K_2(x))\ =\ -K_0(x)-\frac1xK_1(x).
\label{difrec}
\end{eqnarray}
After some simple algebra for the most tedious case of four functions 
$K_1(x)$ in the integrand we obtain the recursion relation
\begin{eqnarray}
\lefteqn{\intl K_1^4(x)x^qdx
  \ =\ -\intl\frac{d}{dx}(K_0(x))K_1(x)^3x^qdx}\nonumber\\
  &=&-\Big[K_0(x)K_1^3(x)x^q\Big]_0^\infty\nonumber\\&&
  -3\intl K_0^2(x)K_1^2(x)x^qdx
  +(q-3)\intl K_0(x)K_1^3(x)x^{q-1}dx.\nonumber
\end{eqnarray}
The other cases are simpler and will not be written down here. For $q>m$,
the surface terms of the form  $[K_0^{(4-m)}(x)K_1^m(x)x^q]_0^\infty$ vanish.
Therefore the only elements of this recursion are the integrals $I_m(q)$,
and the recursion is expressed as
\begin{eqnarray}
I_4(q)&=&(q-3)I_3(q-1)-3I_2(q),\nonumber \\
I_3(q)&=&\frac12((q-2)I_2(q-1)-2I_1(q)),\nonumber\\
I_2(q)&=&\frac13((q-1)I_1(q-1)-I_0(q)),\nonumber\\
I_1(q)&=&\frac14qI_0(q-1),
\label{zrec}
\end{eqnarray}
which reduces the starting integrals to our basis integrals
\begin{equation}
I_0(q)=\intl K^4_0(x)x^qdx.
\label{i0bas}
\end{equation}

\subsection{Reduction to integrals containing $K_0^4(x)\ln x$}
We again use the relations~(\ref{difrec}) to reduce the general water melon
integrals to those containing only the McDonald function $K_0(x)$. This is
done again by partial integration. As in the previous subsection we find 
\begin{eqnarray}
\lefteqn{\intl K_1^4(x)x^q\ln x\,dx
  \ =\ -\intl K_1^3(x)\frac{dK_0(x)}{dx}x^q\ln x\,dx}\nonumber\\
  &=&-\Big[K_0(x)K_1^3(x)x^q\ln x\Big]_0^\infty
  -3\intl K_0^2(x)K_1^2(x)x^q\ln x\,dx\nonumber\\&&
  +(q-3)\intl K_0(x)K_1^3(x)x^{q-1}\ln x\,dx
  +\intl K_0(x)K_1^3(x)x^{q-1}dx\nonumber 
\end{eqnarray}
for the case with four functions $K_1(x)$ in the integrand. For integer
$q>m$, the surface terms $[K_0^{4-m}(x)K_1^m(x)x^q\ln x]_0^\infty$ again
vanish. Therefore the recursion is expressed in terms of the integrals
$I_m^l(q)$ and is given by
\begin{eqnarray}
I_4^l(q)&=&(q-3)I_3^l(q-1)+I_3(q-1)-3I_2^l(q),\nonumber\\[3pt]
I_3^l(q)&=&\frac12((q-2)I_2^l(q-1)+I_2(q-1)-2I_1^l(q)),\nonumber\\
I_2^l(q)&=&\frac13((q-1)I_1^l(q-1)+I_1(q-1)-I_0^l(q)),\nonumber\\
I_1^l(q)&=&\frac14(qI_0^l(q-1)+I_0(q-1)).
\label{logrec}
\end{eqnarray}
Together with Eqs.~(\ref{zrec}) these relations give the complete set of
one-parameter recurrence equations for reducing a general water melon
integral to a set of master integrals of the form $I_0(q)$ from
Eq.~(\ref{i0bas}) and 
\begin{equation}
I_0^l(q)=\intl K^4_0(x)x^q\ln x\,dx.
\label{ilbas}
\end{equation}
Concluding these two subsections we affirm that Eqs.~(\ref{i0bas})
and~(\ref{ilbas}) constitute our basis for the evaluation of water melon
diagrams with high powers of denominators.

\subsection{Efficiency of reduction}
Note that the direct reduction of a water melon diagram to the master
integrals is rather slow within the straightforward application of
recurrence relations based on momentum space representation. In practice
the computation proceeds through the use of a table of integrals with given
powers of the denominators. One would have a three-dimensional table for a
given total power $N$ if no modifications of the basic technique as
developed in ref.~\cite{b3b4} have been introduced. The number of entries
(even when accounting for the appropriate symmetries) then grows as fast as
$N^3$ which is large for the large values of $N$ needed in some present
applications. Within our methods one first re-expresses these integrals
through a one-parameter set of integrals which are solved explicitly. For
large $N$ the number of entries increases as a first power of $N$ (the
number of elements for the $I_0(q)$  basis is given by $2[N/2]-5$ where
$[z]$ is an integer part of $z$) which considerably reduces the time
consumption in a computer evaluation. Note that in refs.~\cite{Steinh,BaiSte}
different recursion techniques have been described which also avoid the use
of the three-dimensional tables in reducing the water-melon diagram. For
instance, in the package MATAD~\cite{Steinh} the three-loop water melon
diagrams are reduced to a one-dimensional table of integrals using a
dedicated set of (momentum-based) recurrence relations.

\section{Computation}
In this section we give some explicit formulae for computing water melon
diagrams within the configuration space technique. The methods is best used
for direct numerical computation or for constructing efficient approximation
formulae for water melon diagrams. Therefore, as for exact analytical
expressions, we mainly consider the $D=4$ case where the initial values of
the recurrence procedure can be easily found and can be compared
with numerical results. The construction of the $\eps$-expansion is not
drastically simplified as compared to the standard momentum space approach
and we therefore do not dwell on it here.
 
\subsection{$N=4$ water melon}
We describe a simple way of computing an initial value of a general $D=4$
(or, more precisely, integer dimension) water melon diagram. The quantity we
need has the form
\begin{equation}
I(q)=\intl K^4_0(x)x^{2q+1}dx\equiv I_0(2q+1)
\label{2i}
\end{equation}
and represents a member of our basis.

We concentrate first on the case $q=0$. Note that Eq.~(\ref{2i}) is simply
the result for a water melon diagram with a massive propagator $K_0(mx)$
within a two-dimensional theory~\cite{schPL}. The corresponding two-line
water melon (master one-loop diagram) in momentum space has the explicit
form
\begin{equation}
\tilde \Pi_2(p)=\frac1{2\pi\sqrt{p^2}\sqrt{p^2+4m^2}}
\ln\left(\frac{\sqrt{p^2+4m^2}+\sqrt{p^2}}{\sqrt{p^2+4m^2}-\sqrt{p^2}}\right).
\label{2wmp}
\end{equation}
Eq.~(\ref{2i}) now becomes
\begin{equation}
I(0)=2\pi m^2\int\tilde\Pi_2(p)^2d^2p=\frac14\intl\frac{\xi^2d\xi}{\sinh\xi}
  =\frac78\zeta(3)
\label{2final}
\end{equation}
where we have changed the integration variable $p$ to a new variable $\xi$
defined by the relation $p=2m\sinh(\xi/2)$ and used the standard
integral~\cite{Brychkov}
\begin{equation}
\intl\frac{\xi^{\alpha-1}d\xi}{\sinh\xi}
  =\frac{2^{\alpha}-1}{2^{\alpha-1}}\Gamma(\alpha)\zeta(\alpha)
\label{tabint}
\end{equation}
with $\Gamma(\alpha)$ being Euler's $\Gamma$-function and $\zeta(\alpha)$
being Riemann's $\zeta$-function. Note that the two-dimensional one-loop
correlator in momentum space $\tilde\Pi_2$ shown in Eq.~(\ref{2wmp})
has the simple form
\[
\tilde \Pi_2(2m\sinh(\xi/2))=\frac1{4\pi m^2}\frac\xi{\sinh\xi}
\]
when expressed in terms of the new variable $\xi$. 
The integration measure becomes $d^2p=4\pi m^2\sinh\xi\,d\xi$.

Results for other values of $q$ can be obtained by differentiating
one of the two $\tilde\Pi_2$ in the integrand in Eq.~(\ref{2final}),
\begin{equation}
I(q)=2\pi m^2\int\tilde\Pi_2(p)(- m^2\dalembertian_p)^q\tilde\Pi_2(p)d^2p
\label{2qgen}
\end{equation}
where $\dalembertian_p$ is a two-dimensional d'Alembert operator in (Euclidean)
momentum space, $\dalembertian_p=\partial^2/\partial p_\mu\partial p_\mu$.
There is a possibility to differentiate a separate line of this three-loop
water melon diagram that leads to different representations for higher moments
but we find Eq.~(\ref{2qgen}) to be the most convenient one. We do not have
a general analytical solution to Eq.~(\ref{2qgen}) for arbitrary large $q$
at the moment, although the solutions for some first values of $q$ are
easily available.

Also we comment on the general structure of the basis given by $I(q)$ in
Eq.~(\ref{2i}). This basis has the form 
\begin{equation}
  \label{zbasex}
A_q\zeta(3)-B_q  
\end{equation}
where the
transcendental number $\zeta(3)$ is manifestly written down while $A_q$,
$B_q$ are rational positive numbers for any $q$. Several pairs of
coefficients $(A_q,B_q)$ for $q=1,2,3,4,5$ are 
\[
\left(\frac{7}{32},\frac{3}{16}\right),
\quad \left(\frac{49}{128},\frac{27}{64}\right),
\quad \left(\frac{63}{32},\frac{37}{16}\right),
\quad \left(\frac{42777}{2048},\frac{25555}{1024}\right),
\quad \left(\frac{3101175}{8192},\frac{9304913}{20480}\right).
\]
The quantity $I(q)$ is positive for any $q$. Its numerical magnitude can
be easily inferred from the asymptotic expansion of the integral given in
Eq.~(\ref{2i}) at large $q$,
\begin{equation}
  \label{asym}
I(q)=\frac{\pi^2\Gamma(2q)}{4^{2q+1}}\left(1-\frac{1}{q-1/2}+O(1/q^2)\right).
\end{equation}
However, its analytical representation given in Eq.~(\ref{zbasex}) reveals
a rather awkward behaviour when analyzed numerically. Two terms with
manifestly different transcendental structure taken as analytical
expressions cancel each other to a large extent upon numerical evaluation. 
Indeed, using the numerical value of the transcendental number
$\zeta(3)=1.20206\ldots\,$, the numerical values of the pairs
$(A_q\zeta(3),B_q)$ for $q=1,2,3,4,5$ are given by
\begin{eqnarray}
&&(0.26295,0.1875),\quad(0.460162,0.421875),\quad(2.36655,2.3125),\nonumber\\
&&\qquad\qquad(25.1076,24.9561),\quad(455.052,454.341).\nonumber
\end{eqnarray}
For $q=5$ the numerical values of quantities of different transcendentality
coincide with each other up to three significant figures resulting in a
cancellation on subtraction and a large loss of precision. For larger $q$
the cancellation of the first significant figures of the two numbers is
more dramatic resulting in even larger losses of numerical precision. It
casts some doubt on whether the presentation of results in terms of their
manifest transcendental structure is the most preferable one. On the other
hand the asymptotic formula Eq.~(\ref{asym}) is simple and quite precise.
Namely, taking Eq.~(\ref{asym}) and introducing a new parameter $\kappa$
which accounts for higher order terms in the asymptotic expansion, one has
\begin{equation}
  \label{asym1}
I(q)=\frac{\pi^2\Gamma(2q)}{4^{2q+1}}\left(1-\frac1{q+\kappa}\right).
\end{equation}
For $\kappa= 0.97$ this simple formula gives good numerical results for the
basis integrals with an accuracy better than $1\%$ for all $q\ge 1$. For
$q>3$ the relative accuracy is better than $10^{-3}$ (one per mille).

Another representation for the basis set of integrals can be obtained using
the dispersion relation in configuration space. One obtains the relation 
\begin{equation}
K_0^4(x)=(2\pi)^3\int_{16}^\infty\rho_4(s)K_0(\sqrt sx)ds
\label{sdisp}
\end{equation}
with $\rho_4(s)$ being a spectral density for the three-loop water melon
diagram. Note that such a representation is a basis for the sum rule
calculations in configuration space both in two-dimensional~\cite{schPL} and
four-dimensional space-time~\cite{xChetPiv}. One obtains the following
relation for the basis integrals in terms of the moments of the spectral
density $\rho_4(s)$,
\begin{equation}
I(q)=(2\pi)^3 4^{q}(q!)^2\int_{16}^\infty {\rho_4(s)\over s^{q+1}}ds.
\label{disprep}
\end{equation}
An efficient way of computing the spectral density of water melon diagrams
has recently been developed~\cite{xSpectr}. However, the direct
configuration space representation is the most convenient for a numerical
evaluation.  To emphasize this last remark, we give an asymptotic formula
for the $\eps$-expansion of a water melon diagram where $\ln(x)$ appears
in the integrand. The asymptotic formula for the log-type integrals reads
(where we changed our notation by introducing $\ln(4x)$ for convenience)
\begin{eqnarray}
L^l_4(q)&=&\intl K^4_0(x)x^{2q+1}\ln(4x)dx\nonumber\\
  &=&\frac{\pi^2\Gamma(2q)}{4^{2q+1}}\Psi(2q)
  \left(1-\frac{1}{q-1/2}+O(1/q^2)\right)
\label{asymlog}
\end{eqnarray}
where $\Psi(x)=\Gamma'(x)/\Gamma(x)$ is the logarithmic derivative of the
$\Gamma$-function.

Modifying the last term by introducing a parameter $\kappa_l$ as before in
Eq.~(\ref{asym1}) we find that the relation 
\begin{equation}
L^l_4(q)
  =\frac{\pi^2\Gamma(2q)}{4^{2q+1}}\Psi(2q)\left(1-\frac1{q+\kappa_l}\right)
\label{kapasymlog}
\end{equation}
with $\kappa_l=1.17$ gives good results for the basis log-type integrals
with an accuracy better than $1\%$ for all $q>3$. For $q>5$ the relative
accuracy is better than $10^{-3}$. The leading order asymptotic formula 
presented in Eq.~(\ref{kapasymlog}) is less precise for small $q$ than its
analog in Eq.~(\ref{asym1}) in the previous case because the integrand in
Eq.~(\ref{asymlog}) is not positive for log-type integrals. It is also
clear that an exact solution of the recurrence relations will be much more
complicated than these simple asymptotic formulae. For large $q$ the
numerical cancellation of terms with different transcendental structure up
to very high significant figures will be quite dramatic.

The solution of the recurrence relations depends on the space-time
dimensionality. In the case $D=3$ the structure of the recurrence relations
drastically simplifies. The basis set of integrals in three-dimensional
space-time can be written down explicitly. The propagator is 
\begin{equation}
D(x,m)=\frac1{4\pi x}e^{-mx}
\label{D3prop}
\end{equation}
which leads to an integral basis of the form
\begin{equation}
  H_0(q)=(2\pi)^{4(\lambda+1)}\intl D(x,m)^4x^{q+2}dx\rightarrow
  \frac{\pi^2}4\intl e^{-4x}x^{q-2}dx=\frac{\pi^2}{4^q}\Gamma(q-1).
\label{b3bas}
\end{equation}
The special case $D=3$ can serve as a check of any general solution
of the recurrence relations.

\subsection{$N=3$ water melon: a standard sunset}
The case of a two-loop water melon (genuine sunset) is simple indeed and
can be easily analyzed along the same lines. The corresponding basis set of
configuration space integrals is quite analogous to the previous case and
is simpler because it now includes only three McDonald functions,
\begin{eqnarray}
J_n(q)&=&\intl K_0^{3-n}(x)K_1^n(x)x^qdx,\nonumber\\
J_n^l(q)&=&\intl K_0^{3-n}(x)K_1^n(x)x^q\ln x\,dx.
\label{defbas3}
\end{eqnarray}
The reduction to the basis set of integrals analogous to the case of
three-loop water melons given in Eqs.~(\ref{i0bas}) and~(\ref{logrec}) can
be now readily obtained.

The basic initial integral (the basic sunset $B_S$) for the recurrence
relation has the explicit form 
\begin{equation}
B_S=\int\frac{\tilde \Pi_2(p)}{p^2+M^2}d^2p
\label{3linesp}
\end{equation}
with $\tilde \Pi_2(p)$ given by Eq.~(\ref{2wmp}). Here $M$ is a mass of
the third line which is kept different from the other two with masses $m$.
By differentiating with respect to $M$, any positive power of propagator
(and/or power of $x$ in configuration space representation) can be
obtained. After changing the integration variable as in the preceding
subsection one finds an explicit representation
\begin{equation}
B_S=\intl\frac{\xi d\xi}{4m^2\sinh^2(\xi/2)+M^2}.
\label{3lines}
\end{equation}
After a change of variables the integration can be done and can be reduced
to a polylogarithm function. Namely, for $t=e^{-\xi}$ one has
\begin{equation}
B_S=\intl\frac{\xi\,d\xi}{4m^2\sinh^2(\xi/2)+M^2}=
  -\frac1{m^2}\int_0^1\frac{\ln t\,dt}{1-2\gamma t+t^2}
\label{3lines00}
\end{equation}
where $\gamma=1-M^2/2m^2$. The last integral is evaluated to give
\begin{equation}
\int_0^1\frac{\ln t\,dt}{1-2\gamma t+t^2}
  =\frac{{\rm Li}_2(1/t_1)-{\rm Li}_2(1/t_2)}{t_1-t_2}
\label{3lines000}
\end{equation}
where $t_{1,2}=\gamma\pm\sqrt{\gamma^2-1}$ and ${\rm Li}_2(z)$ is the
dilogarithm function, 
\[
{\rm Li}_2(z)=\sum_{k=1}^\infty\frac{z^k}{k^2},\qquad|z|<1.
\]
The differentiation with respect to $M$ is now straightforward and can be
performed with a symbolic manipulation program.
 
In the case $M=2m$ the integration simplifies because the two independent
parameters $M$ and $2m$ on which the integrand depends, coincide
(degenerate case). The integral is then reduced to a special case of
Eq.~(\ref{tabint}). We have $\gamma=-1$ and 
\begin{equation}
\int_0^1\frac{\ln t\,dt}{(1+t)^2}=-\ln 2 
\label{g-1}
\end{equation}
which leads to 
\begin{equation}
B_S(M=2m)=\frac{\ln 2}{m^2}.
\label{M2msun}
\end{equation}
For the case $M=m$ the standard result -- Clausen's polylogarithm
${\rm Cl}_2(\pi/3)$ -- is reproduced (see e.g.\ ref.~\cite{broad}). Indeed,
$\gamma=1/2$ and $t_{1,2}=\exp({\pm i\pi/3})$. Eq.~(\ref{3lines000}) now
becomes
\begin{equation}
\int_0^1\frac{\ln t\,dt}{1-t+t^2}
  =-\frac2{\sqrt3}{\rm Im}\,{\rm Li}_2(e^{i\pi/3})
\label{imcl}
\end{equation}
and with using the definition of Clausen's polylogarithm,
${\rm Cl}_2(\theta)={\rm Im}\,{\rm Li}_2(e^{i\theta})$
one finds
\begin{equation}
B_S(M=m)=\frac2{m^2\sqrt3}{\rm Cl}_2\left(\frac\pi3\right).
\label{M1msun}
\end{equation}

\subsection{Generalization to the spectacle topology}
In this subsection we give a formula for a more general topology when only
one propagator is removed from the initial three-loop bubble diagram. In
the original classification of ref.~\cite{avdeev} these are class $E$
diagrams belonging to the spectacle topology. The formula obtained in
this subsection is efficient for numerical integration though we did not
find any analytical solution. The main obstacle of generalizing  
the configuration space technique to a general multi-loop diagram is the 
angular integration. The configuration space technique proved to be rather
successful for general diagrams in the massless case~\cite{Geg} but it
brings no essential simplification in the general massive case (see e.g.\
ref.~\cite{mendels}). However, for special configurations the angular
integration can be explicitly performed with a reasonably simple integrand
left for the radial integration. The diagrams of spectacle topology give
an example of such a configuration.

The configuration space expression of a spectacle topology diagram written
in a form suitable for our purpose is (see Fig.~2(a))
\begin{equation}
\int D(x-y,m)D(x,m)^2d^DxD(y,m)^2d^Dy.
\label{spectxexp}
\end{equation}
The key relation for a drastic simplification of the configuration space
integral with the spectacle topology is the addition theorem for Bessel
functions allowing one to perform some angular integration explicitly. One
needs to integrate over the relative angle in the propagator $D(x-y,m)$.
In the handbook of Gradshteyn and Rhyshik~\cite{GrRy} one finds
\begin{equation}
\frac{Z_\nu(mR)}{R^\nu}=2^\nu m^{-\nu}\Gamma(\nu)
  \sum_{k=0}^\infty(\nu+k)\frac{J_{\nu+k}(m\rho)}{\rho^\nu}
  \frac{Z_{\nu+k}(mr)}{r^\nu}C^\nu_k(\cos\varphi)
\label{addth}
\end{equation}
where $C^\nu_k$ are the Gegenbauer polynomials, $Z$ is any of 
the Bessel functions $J$, $Y$, $H^{(1)}$ or $H^{(2)}$,
\begin{equation}
R=\sqrt{r^2+\rho^2-2r\rho\cos\varphi}
\end{equation}
and $r>\rho$. For $r<\rho$ the arguments of the Bessel functions on the
right hand side of Eq.~(\ref{addth}) should be interchanged. Writing
$R=|r-\rho|$ we have for $r>\rho$ 
\begin{equation}
\frac{Z_\nu(m|r-\rho|)}{|r-\rho|^\nu}=2^\nu m^{-\nu}\Gamma(\nu)
  \sum_{k=0}^\infty(\nu+k)\frac{J_{\nu+k}(m\rho)}{\rho^\nu}
  \frac{Z_{\nu+k}(mr)}{r^\nu}C^\nu_k(\cos\varphi) .
\end{equation}
Using this relation for $Z=H^{(1)}$ and substituting $m=e^{i\pi/2}$ for the
purpose of analytic continuation in order to obtain a relation for the
modified Bessel functions $K$ and $I$, we find
\begin{eqnarray}
\frac{K_\lambda(|r-\rho|)}{|r-\rho|^\lambda}
  &=&2^\lambda\Gamma(\lambda)\sum_{k=0}^\infty(\lambda+k)
  \frac{I_{\lambda+k}(\rho)}{\rho^\lambda}
  \frac{K_{\lambda+k}(r)}{r^\lambda}C_k^\lambda(\cos \varphi)
\label{eqn1}
\end{eqnarray}
where we have
\begin{eqnarray}
K_\lambda(z)&=&\frac{i\pi}2e^{\pi\lambda i/2}H_\lambda^{(1)}(iz),\nonumber\\
I_\lambda(z)&=&e^{-\pi\lambda i/2}J_\lambda(e^{\pi i/2}z)\qquad
  \mbox{for\ }-\pi<\mbox{arg}\,z\le\frac\pi2,\nonumber\\
I_\lambda(z)&=&e^{3\pi\lambda i/2}J_\lambda(e^{-3\pi i/2}z)\qquad
  \mbox{for\ }\frac\pi2<\mbox{arg}\,z\le\pi.
\end{eqnarray}

Using the orthogonality relations for Gegenbauer polynomials (see the
Appen\-dix), the sum disappears after integration over the relative angle
and only one term contributes. We obtain
\begin{eqnarray}
\int\frac{K_\lambda(|r-\rho|)}{|r-\rho|^\lambda}d\Omega_\rho
  &=&2^\lambda\Gamma(\lambda)\sum_{k=0}^\infty(\lambda+k)
  \frac{I_{\lambda+k}(\rho)}{\rho^\lambda}\frac{K_{\lambda+k}(r)}{r^\lambda}
  \int C_k^\lambda(\cos\varphi)d\Omega_\varphi \nonumber\\
  &=&2^\lambda\Gamma(\lambda)\lambda
  \frac{I_{\lambda}(\rho)}{\rho^\lambda}\frac{K_{\lambda}(r)}{r^\lambda}
  \frac{2\pi^{\lambda+1}}{\Gamma(\lambda+1)}C_0^\lambda(1)
  \nonumber\\
  &=& (2\pi)^{\lambda+1}\frac{I_\lambda(\rho)}{\rho^\lambda}
  \frac{K_\lambda(r)}{r^\lambda}, \qquad {r>\rho},
\end{eqnarray}
where the first equality is a consequence of the orthogonality relation
with the trivial factor $C_0^\lambda(1)=1$. This result allows one to write
down an expression for any spectacle-type diagram in the form of a two-fold
integral with a simple integration measure
\begin{eqnarray}
&&\intl D(x,m)^2x^{2\lambda+1}dx\intl D(y,m)^2y^{2\lambda+1}dy
  \ \times\nonumber\\&&\qquad\left(
  \frac{K_\lambda(x)}{x^\lambda}\frac{I_\lambda(y)}{y^\lambda}\theta(x-y)
  +\frac{K_\lambda(y)}{y^\lambda}\frac{I_\lambda(x)}{x^\lambda}\theta(y-x)
  \right)
\label{spect}
\end{eqnarray}
where $\theta(x)$ is the standard step-function distribution.

Note that the integration measure $D(x,m)^2x^{2\lambda+1}dx$ allows one to
perform the integration by using efficient integration routines for a
numerical evaluation. The form of the weight function is close to
$e^{-ax}x^{\alpha}$  which makes the use of Laguerre polynomials a
convenient choice within the Gaussian numerical integration method. Any
modified propagator (with any power of the denominator) can be used as a
factor in the integration measure $D(x,m)^2x^{2\lambda+1}dx$ which makes
this representation universal and useful for the case of high powers of
denominators of the lines associated with pairs $(x,0)$ and $(y,0)$ of
space-time points. If the angular structure of the diagram is preserved,
the generalization to higher loops in the expressions for the radial
measures is straightforward. 

As an illustration of this technique we present an example of the evaluation
of a spectacle diagram. Consider an integer dimension space-time which,
without loss of generality, can be chosen to be two-dimensional (an
odd number of dimensions is trivial because the propagators degenerate
to simple exponentials). The spectacle-type three-loop diagram can be
obtained in a closed form. Indeed, in the momentum space representation we
have
\begin{equation}
S(M)=\int\frac{{\tilde \Pi_2(p)}^2}{p^2+M^2}d^2p  
\label{sp1}
\end{equation}
for the basic spectacle diagram $S$ with $\tilde\Pi_2(p)$ taken from
Eq.~(\ref{2wmp}) and the mass $M$ of the connecting propagator kept
different. After the substitutions $p=2m\sinh(\xi/2)$ and $t=e^{-\xi}$ we
have
\begin{equation}
S(M)=\frac1{2\pi m^4}\int_0^1\frac{t\,\ln^2t\,dt}{(1-t^2)(1-2\gamma t+t^2)}.
\label{spcha}
\end{equation}
Performing the integration we finally obtain
\begin{equation}
S(M)=\frac{f(t_1)-f(t_2)}{t_1-t_2}
\label{manint}
\end{equation}
with
\[
f(t)=\frac{8t{\rm Li}_3(1/t)-(t+7)\zeta(3)}{8\pi m^4(t^2-1)}
\]
where $t_{1,2}=\gamma\pm\sqrt{\gamma^2-1}$ with $\gamma=1-M^2/2m^2$ as
before. ${\rm Li}_3(z)$ is the trilogarithm function
\[
{\rm Li}_3(z)=\sum_{k=1}^\infty\frac{z^k}{k^3}, \quad |z|<1.
\]
For $M=2m$ the integral in Eq.~(\ref{spcha}) simplifies as in the case
of the sunset diagram and one finds a simple answer in terms of the standard
(in the present context) transcendental numbers $\ln 2$ and $\zeta(3)$,
\[
S(2m)=\frac1{4\pi m^4}\left(\frac78\zeta(3)-\ln 2\right).
\]
For the actual value of the mass $M=m$ we obtain a result including the
next Clausen polylogarithm ${\rm Cl}_3(2\pi/3)$. As one can conclude
from this expression, the conjugate pair of the sixth order roots
of unity, $\exp(\pm 2\pi i/3)$ plays an important role in this case again
in accordance with the general analysis of ref.~\cite{broad}. The
origin of the appearance of the the sixth order roots of unity as the
parameters of the analytical expressions of the diagrams lies in the
mismatch of masses along the lines of the diagrams. However, the
exceptional case $M=2m$, where one line has the double mass of the other
lines (which results in the drastic simplification) also keeps us within
the set of the sixth order roots of unity. The key parameter in this case
is simply the natural  number $1$ which definitely is one of the sixth
order roots of unity.

Turning to the configuration space representation we find
\begin{eqnarray}
S(M)&=&\intl xK_0(mx)^2dx\intl yK_0(my)^2dy\ \times \nonumber\\
&&\Bigg(K_0(Mx)I_0(My)\theta(x-y)+K_0(My)I_0(Mx)\theta(y-x)\Bigg)
\label{xMc00}
\end{eqnarray}
for the basic spectacle diagram. An explicit numerical integration of
Eq.~(\ref{xMc00}) shows coincidence with the analytical result in
Eq.~(\ref{manint}) which we checked for arbitrary values of $M$. In
this example the analytical result has a rather simple form which is not
true if high powers of denominators enter. Then the corresponding one-loop
insertions are rather cumbersome and an explicit integration in
configuration space is more convenient.

As a last topic of this subsection we demonstrate how the square of 
Clausen's polylogarithm ${\rm Cl}_2(\pi/3)^2$ can emerge at the level of
spectacle topology diagrams. The transcendental number ${\rm Cl}_2(\pi/3)^2$
characterizes the analytical results for three-loop bubbles. Its presence
was discovered in the impressive treatise of David Broadhurst on the role
of the sixth order roots of unity for the transcendentality structure of
results for Feynman diagrams in quantum field theory~\cite{broad}.

We consider the spectacle diagram in the form shown in Fig.~2(b). Here the
expression for the generalized middle line is a product of the one-loop
propagator of Eq.~(\ref{2wmp}) and the standard particle propagator. We
express the one-loop propagator in Eq.~(\ref{2wmp}) using a dispersion
representation with the spectral density $\rho_2(s)$ and obtain
\begin{eqnarray}
&&{\rm Loop}(p,m)\times{\rm Propagator}(p,m)
=\tilde \Pi_2(p)\frac1{p^2+m^2}  
=\frac1{p^2+m^2}\int_{4m^2}^\infty\frac{\rho_2(s)ds}{s+p^2}\nonumber\\
  &&\qquad=\int_{4m^2}^\infty\frac{\rho_2(s)ds}{s-m^2}
  \left(\frac1{p^2+m^2}-\frac1{s+p^2}\right).
\label{midline}
\end{eqnarray}
Taking only the first term (which is sufficient for obtaining the
result we are looking for) one has 
\begin{equation}
I=\frac1{p^2+m^2}\int_{4m^2}^\infty\frac{\rho_2(s)ds}{s-m^2}
\label{ii}
\end{equation}
which leads to the sunset diagram after the two other line shown in
Fig.~2(b) have been added with a normalization factor given by the integral.
One factor ${\rm Cl}_2(\pi/3)$ results from integrating the overall sunset
diagram which is composed of the propagator $(p^2+m^2)^{-1}$ from
Eq.~(\ref{ii}) with the two other lines of the diagram shown in Fig.~2(b).
The second factor ${\rm Cl}_2(\pi/3)$ has to be found in the normalization
factor given by the integral in Eq.~(\ref{ii}). Note that the very
structure of this contribution -- the square of a number which first
appeared at the lower loop level -- suggests a hint for its search. It
should emerge as an iteration of a lower order contribution in accordance
with the iterative structure of the $R$-operation (see e.g.\ ref.~\cite{Rop})
which forms a general framework for the analysis of multiloop diagrams.
The following consideration confirms this guess. Consider the quantity
\[
N=\int_{4m^2}^\infty {\rho_2(s)ds\over s-m^2}
\]
and take $\rho_2(s)$ to be the spectral density in $D$-dimensional
space-time (see e.g.\ Eq.~(65) of ref.~\cite{wm}),
\begin{equation}
\rho_2(s)=\frac{(s-4m^2)^{\lambda-1/2}}
  {2^{4\lambda+1}\pi^{\lambda+1/2}\Gamma(\lambda+1/2)\sqrt s},\qquad\sqrt s>2m.
\end{equation}
Now consider a first order contribution of the expansion in $\eps$
near the space-time dimension $D=2$. The expansion in $\lambda=-\eps$
results in
\begin{equation}
\frac{(s-4m^2)^{-\eps-1/2}}{\mu^{2\eps}\sqrt s}=\frac1{\sqrt{s(s-4m^2)}}
  \left(1-\eps\ln\left(\frac{s-4m^2}{\mu^2}\right)+O(\eps^2)\right),
\end{equation}
so the relevant first order term in $\eps$ is
\begin{equation}
\Delta_\eps\rho_2(s)=-\frac{\ln((s-4m^2)/m^2)}{2\pi\sqrt{s(s-4m^2)}}
\end{equation}
where $\mu=m$ has been chosen for convenience. Now we change the variable
according to
\begin{equation}
\sqrt s=2m\cosh(\xi/2),\qquad t=e^{-\xi}
\end{equation}
to obtain
\begin{equation}
\Delta_\eps\rho_2(4m^2\cosh^2(\xi/2))
  =\frac{(\ln t-2\ln(1-t))t}{2\pi m^2(1-t^2)}.
\end{equation}
For the quantity in question we find
\begin{equation}
\Delta_\eps N=\int_{4m^2}^\infty\frac{\Delta_\eps\rho_2(s)ds}{s-m^2}
  =\frac1{2\pi m^2}\int_0^1\frac{(\ln t-2\ln(1-t))dt}{1+t+t^2}.
\label{delnorm}
\end{equation}
The roots of the denominator of the integrand in Eq.~(\ref{delnorm})
are now $t_{3,4}=\exp({\pm 2\pi i/3})$ which again is a conjugate pair of
the sixth order roots of unity. After integrating this equation we readily
find
\begin{eqnarray}
\Delta_\eps N&=&\frac1{2\pi m^2\sqrt3}\left(
  {\rm Im}\left(\Li_2\left(e^{2i\pi/3}\right)
  -\Li_2\left(e^{-2i\pi/3}\right)\right)-\frac\pi3\ln3\right)\nonumber\\
  &=&-\frac1{\pi m^2\sqrt 3}\left({\rm Cl}_2\left(\frac{2\pi}3\right)
  -\frac\pi6\ln3\right).
\end{eqnarray}
Using the relation
\[
{\rm Cl}_2\left(\frac{2\pi}3\right)=\frac23{\rm Cl}_2\left(\frac\pi3\right)
\]
we finally obtain
\begin{equation}
\Delta_\eps N=\frac2{3\pi m^2\sqrt3}
  \left({\rm Cl}_2\left(\frac\pi3\right)-\frac\pi4\ln3\right).
\end{equation}
Therefore, in the first order of the $\eps$ expansion of the spectacle
diagram we found this remarkable contribution proportional to
${\rm Cl}_2(\pi/3)^2$.  In our calculation it emerges naturally as the
iteration of the lower order term. Originally this contribution had been
guessed and confirmed in ref.~\cite{broad} by a direct numerical computation
of the finite part of the general three-loop bubble in four-dimensional
space-time.  

Note also that besides $B_4$ and ${\rm Cl}_2(\pi/3)^2$ there is one more
nontrivial transcendental number necessary for the analytical computation
of the finite part of the general three-loop bubble in four-dimensional
space-time as has been proven in ref.~\cite{broad}. This number was
chosen in ref.~\cite{broad} as a two dimensional nested sum of the form 
\[
V_{3,1}=\sum_{m>n>0}\frac{(-1)^m\cos(2\pi n/3)}{m^3 n}.
\]
The analogous contribution may also appear in the first order
$\eps$-expansion of the spectacle diagram. We hope to devote a separate
publication to this point.

Therefore there are solid arguments that all basic transcendental numbers
necessary for the analytical computation of three-loop massive bubbles can
be found at the level of much simpler topology than the general tetrahedron.
One may need only spectacle diagrams as the largest set to find all
necessary transcendental numbers. These discoveries lead us to a conjecture
about the extension of recurrence relations for three-loop bubbles beyond
those found within the integration by part technique. Indeed, if all
results can be obtained within the spectacle topology only, it looks
plausible that there is a procedure that can perform a reduction of the
general tetrahedron diagram to a simpler set where the spectacle topology
is the most complicated one. The manifest form of this procedure is not
known at present.

\section{Conclusion}
We have formulated a new representation for some of the massive three-loop
vacuum diagrams with the simple water melon topology. We have obtained
numerical values for some master integrals within our representation. The
computation of diagrams with large powers of propagators is reduced to a
linear (one-parameter) set of basis integrals. Our representation is simple
and provides a tool for an efficient numerical evaluation of such diagrams.
We have given asymptotic estimates which provide an accuracy better than one
per mille for all basis integrals with $q>3$ in the case of the leading
order of $\eps$-expansion and with $q>5$ in the case of the first order of
$\eps$-expansion. The generalization to higher loops is straightforward.
We gave strong arguments that all nontrivial transcendental numbers
necessary for the computation of three-loop bubbles (which were identified
in ref.~\cite{broad} by using a high precision numerical integration) 
already appears at the level of spectacle topology diagrams.

As a by-product of our analysis we obtained analytical results for a
number of definite integrals involving products of four McDonald functions
of different orders with powers and logarithms (Sec.~3). 

\subsection*{Acknowledgments}
We are indebted to K.G.~Chetyrkin and D.~Broadhurst for discussions,
constructive criticism, remarks on the literature and friendly advice.
A.A.~Pivovarov thanks P.A.~Baikov for an illuminating discussion of the
present status of the optimization of recurrence relations under study by 
him. The work is supported in part by the Volkswagen Foundation under
contract No.~I/73611. A.A.~Pivovarov is supported in part by the Russian
Fund for Basic Research under contracts Nos.~97-02-17065 and 99-01-00091.
S.~Groote gratefully acknowledges a grant given by the Max Kade Foundation.

\section*{Appendix}
We give some formulae for dealing with Gegenbauer polynomials in Sec.~5.4
that can be found e.g.\ in refs.~\cite{GrRy,Geg,terrano}. The Gegenbauer
polynomials obey the orthogonality relations
\[
\int C_m^\lambda(\hat x_1\cdot\hat x_2)C_n^\lambda(\hat x_2\cdot\hat x_3)
  d\hat x_2=\frac\lambda{n+\lambda}\delta_{mn}
  C_n^\lambda(\hat x_1\cdot\hat x_3)\qquad(\int d\hat x_2=1)
\]
or, written in another form,
\[
\int C_{j_1}^\lambda(\hat a\cdot\hat b)C_{j_2}^\lambda(\hat b\cdot\hat c)
  d\Omega_b=\delta_{j_1j_2}\frac{2\pi^{\lambda+1}}{\Gamma(\lambda+1)}
  \frac\lambda{j_1+\lambda}C_{j_1}^\lambda(\hat a\cdot\hat c).
\]
In particular, we have $C_0^\lambda(x)=1$, $C_1^\lambda(x)=2\lambda x$ and
\[
(j+1)C_{j+1}^\lambda(x)=2(j+\lambda)xC_j^\lambda(x)
  -(j+2\lambda-1)C_{j-1}^\lambda(x)\nonumber
\]
together with
\[
C_j^\lambda(1)=\frac{\Gamma(j+2\lambda)}{j!\Gamma(2\lambda)}.
\]
The characteristic polynomial is given by
\[
(t^2-2tx+1)^{-\lambda}=\sum_{j=0}^\infty t^jC_j^\lambda(x).
\]

\end{document}